\RequirePackage{fix-cm}
\documentclass[smallcondensed]{svjour3}
\smartqed
\usepackage[T1]{fontenc}
\usepackage{times,graphicx,amsmath,amssymb,amsfonts,enumerate,cite,float,marvosym,geometry,needspace,placeins}
\usepackage[colorlinks,bookmarksopen,bookmarksnumbered,citecolor=blue,linkcolor=blue,urlcolor=blue]{hyperref}
\hypersetup{pdftitle={Novel lump solutions of the modified Kadomtsev--Petviashvili-I equation},pdfauthor={Tianwei Qiu and Zhen Wang}}
\graphicspath{{figures/}}
\allowdisplaybreaks[2]
\newcommand{\panel}[3]{\begin{minipage}{#1\linewidth}\centering\includegraphics[width=\linewidth]{#2}\par\small #3\end{minipage}}
\begin{document}
\begin{sloppypar}
\title{Novel lump solutions of the modified Kadomtsev--Petviashvili-I equation}
\titlerunning{Novel lump solutions of the modified Kadomtsev--Petviashvili-I equation}
\author{Tianwei Qiu \and Zhen Wang}
\authorrunning{T. Qiu and Z. Wang}
\institute{T. Qiu \at School of Mathematical Sciences, Beihang University, Beijing 100191, China\\
\and Z. Wang \Letter \at School of Mathematical Sciences, Beihang University, Beijing 100191, China\\\email{wangzmath@163.com}}
\date{}
\setlength{\headerboxheight}{0pt}
\renewcommand{\makeheadbox}{}
\maketitle

\begin{abstract}
We apply the generalized long-wave limit method with spectral perturbations to construct second-order two- and three-lump solutions and third-order three-, four-, five-, and six-lump solutions of the modified Kadomtsev--Petviashvili-I equation. We determine the long-time positions and amplitudes of these degenerate lumps and show that their anomalous scattering produces asymptotic amplitude corrections that differ from those in the Kadomtsev--Petviashvili-I equation.
\end{abstract}

\section{Introduction}\label{sec:intro}
The Kadomtsev--Petviashvili (KP) equation
\[
 (u_t+6uu_x+u_{xxx})_x+3\sigma^2u_{yy}=0
\]
describes weakly nonlinear, weakly dispersive waves with slow transverse variation \cite{kp1970,kpwater}. The choices $\sigma^2=-1$ and $\sigma^2=1$ correspond to the KP-I and KP-II equations, respectively. The modified Kadomtsev--Petviashvili (mKP) equation is related to the KP equation by a two-dimensional Miura transformation \cite{mkp-ori1,mkp-ori2,mkp-lax,mkp-ist} and models electromagnetic waves in ferromagnetic films \cite{mkp-mag}. Its standard form is
\begin{equation}\label{eq:mkp}
 v_t+v_{xxx}-3\sigma^2\left(\frac12v^2v_x-\partial_x^{-1}v_{yy}+v_x\partial_x^{-1}v_y\right)=0,
\end{equation}
where $(\partial_x^{-1}g)(x,y,t)=\int_{-\infty}^{x}g(x',y,t)\,dx'$. We consider the mKP-I equation, for which $\sigma^2=-1$.

Lumps are rational solutions localized in both spatial directions \cite{ablo-lump} and have been observed in gravity-capillary and optical experiments \cite{lumpexperiment,opticlump-PRL}. Bilinear constructions of lump solutions from positive-definite quadratic forms were given in \cite{ma2015,mazhou2018}, and multi-pole (degenerate) lump solutions were obtained by the inverse scattering transform in \cite{multipole}. Unlike lumps with distinct asymptotic velocities, which undergo normal scattering, degenerate lumps share a common asymptotic velocity and undergo anomalous scattering with sublinear relative motion \cite{JETP1993}; higher-order lump patterns for the KP-I equation were analyzed in \cite{yjkkp}. Similar bilinear constructions have been applied to other multi-dimensional nonlinear wave models, including the extended shallow-water equation \cite{wangshallow}, a generalized Fokas equation \cite{wangfokas}, Bogoyavlensky--Konopelchenko-type equations \cite{yangma2026,ma2026}, and a generalized Korteweg--de Vries equation \cite{mabatwa2026}.

In \cite{qiu-mkp}, we studied lumps and lump chains of the mKP-I equation, including a second-order three-lump solution, and in \cite{qiu-arxiv} we developed a generalized long-wave limit (GLWL) method with spectral perturbations for the KP-I equation. In this letter, we use the GLWL method to construct a second-order two-lump solution and third-order three-, four-, five-, and six-lump solutions of the mKP-I equation. We determine their asymptotic positions and amplitudes as $t\to\pm\infty$ and compare the outgoing ($t\to+\infty$) amplitude corrections with those for the KP-I equation. The remainder of this letter is organized as follows. Section~\ref{sec:preliminaries} outlines the GLWL construction. Sections~\ref{sec:second} and \ref{sec:third} present the lump solutions and their asymptotics. Section~\ref{sec:conclusion} discusses the results.

\section{Preliminaries}\label{sec:preliminaries}
\subsection{Soliton solutions}\label{sec:solitons}

As shown in \cite{qiu-mkp}, Eq.~\eqref{eq:mkp} with $\sigma^2=-1$ admits the bilinear representation
\begin{equation}\label{eq:bilinear}
 \begin{aligned}
 &v=2i\left(\ln\frac{f^*}{f}\right)_x=4\operatorname{Im}\frac{f_x}{f},\\
 &(D_y+iD_x^2)f^*\cdot f=0,\qquad (D_t+D_x^3+3iD_xD_y)f^*\cdot f=0     
 \end{aligned}
\end{equation}
for an auxiliary function $f\ne0$. Here, the asterisk denotes complex conjugation, and $D_x$, $D_y$, and $D_t$ are Hirota's bilinear operators \cite{binlinear}.

The $N$-soliton auxiliary function is \cite{qiu-mkp}
\begin{equation}\label{eq:sum}
 \begin{aligned}
 &f_N=\sum_{\chi_1,\ldots,\chi_N=0,1}\exp\left(\sum_{j=1}^N\chi_j\eta_j+\sum_{1\le j<l\le N}\chi_j\chi_l\ln a_{jl}\right),\\
 &\eta_j=k_jx+p_jy+\left(\frac{3p_j^2}{k_j}-k_j^3\right)t+i\arctan\frac{k_j^2}{p_j}+\eta_{0,j},\\
 &a_{jl}=\frac{k_j^2k_l^2(k_j-k_l)^2+(k_jp_l-k_l p_j)^2}{k_j^2k_l^2(k_j+k_l)^2+(k_jp_l-k_l p_j)^2}.
 \end{aligned}
\end{equation}
For the properties of the fundamental lump, we refer the reader to Proposition~1 in \cite{qiu-mkp}.

\subsection{The generalized long-wave limit method with spectral perturbations}\label{sec:construction}

Following our GLWL parameterization \cite{qiu-arxiv}, we set $N=2M$ and
\[
 \begin{aligned}
 &k_j=\beta_j\varepsilon,\qquad p_j=\lambda_jk_j,\qquad
 \lambda_j=\lambda+id_j\varepsilon,\qquad \lambda=a+bi,\quad b>0,\\
 &e^{\eta_{0,j}}=C_{j0}\exp\left(\sum_{s\ge2}C_{js}\varepsilon^s\right),\qquad \beta_{j+M}=\beta_j^*,\qquad \lambda_{j+M}=\lambda_j^*,\qquad
 e^{\eta_{0,j+M}}=(e^{\eta_{0,j}})^*,\qquad 1\le j\le M.
 \end{aligned}
\]
Here, $\beta_j$, $d_j$, and $C_{js}$ are real parameters with $\beta_1=1$, $d_1=0$, and $\beta_jC_{j0}\ne0$. Any higher-order phase parameter $C_{js}$ not specified below is set to zero, and we exclude parameter values for which the interaction coefficients $a_{jl}$ become singular as $\varepsilon\to0$. Expanding $f_{2M}$ in \eqref{eq:sum} in powers of $\varepsilon$ and choosing the parameters to cancel the lower-order terms, we denote by $f_{2M,2n}$ the leading nonzero Taylor coefficient $[\varepsilon^{2n}]f_{2M}$ (up to a nonzero real factor). Substituting $f_{2M,2n}$ into \eqref{eq:bilinear} yields the corresponding lump solution; locally uniform convergence of the normalized auxiliary function and its derivatives away from the zeros of $f_{2M,2n}$ follows as in \cite{qiu-arxiv}.

We refer to $M$ as the order of the GLWL construction. By a peak of $v$ and its amplitude we mean a positive strict local maximum in the $(x,y)$-plane and the value of $v$ there, labeled separately as $t\to-\infty$ and $t\to+\infty$. Throughout the asymptotic analysis, all solution parameters are held fixed, real cube roots are taken, and we assume $a>0$, $b>0$, and that the limiting auxiliary function has no real zeros.

\section{Second-order lump solutions}\label{sec:second}

\subsection{Anomalous scattering of the second-order two-lump solution}\label{sec:two}

For $M=2$, we set $\beta=\beta_2$ and $d=d_2$. Canceling the lower-order Taylor coefficients yields
\[
 \begin{array}{lll}
 \text{Case 1:}&C_{10}=(1+\beta-d)/(1-\beta+d),&C_{20}=(1+\beta+d)/(-1+\beta-d),\\
 \text{Case 2:}&C_{10}=(1+\beta+d)/(1-\beta-d),&C_{20}=(1+\beta-d)/(-1+\beta+d).
 \end{array}
\]
Focusing on Case 1 below, we require $\beta d\ne0$ and
\[
 (1-\beta+d)\bigl((1+\beta)^2-d^2\bigr)\ne0.
\]
The second-order two-lump solution is given by
\begin{equation}\label{eq:two}
 v=4\operatorname{Im}\frac{(f_{4,4})_x}{f_{4,4}},
\end{equation}
where $f_{4,4}$ is given in Appendix \ref{app:functions}. The second-order phases enter only through
\[
 t_0=t-\frac{C_{22}-\beta C_{12}}{6\beta db}.
\]

\begin{proposition}\label{prop:two}
The positions and amplitudes of the two lumps in \eqref{eq:two} satisfy the following asymptotics. As $t\to-\infty$,
\begin{equation}\label{eq:two-in}
 \begin{aligned}
 &x_j=3(a^2+b^2)t+(-1)^j\sqrt{12b|t|}+O(|t|^{-1/2}),\\ 
 &y_j=-6at+O(|t|^{-1/2}),\\
 &A_j=\frac{8b^2}{a}-(-1)^j\frac{8\sqrt b}{\sqrt3\,a}|t|^{-1/2}+O(|t|^{-1}), \quad j=1,2.
 \end{aligned}
\end{equation}
As $t\to+\infty$,
\begin{equation}\label{eq:two-out}
 \begin{aligned}
 &x_j=3(a^2+b^2)t-(-1)^j\frac{\sqrt{12}\,a}{\sqrt b}\sqrt t+\frac1b+O(t^{-1/2}),\\
 &y_j=-6at+(-1)^j\sqrt{12/b}\sqrt t+O(t^{-1/2}),\\
 &A_j=\frac{8b^2}{a}+(-1)^j\frac{4b^{3/2}}{\sqrt3\,a^2}t^{-1/2}+O(t^{-1}), \quad j=1,2.
 \end{aligned}
\end{equation}
\end{proposition}

\begin{proof}
Let $\alpha=a/b$ and $S=1+\alpha^2$, and set
\[
z=b\bigl[x+ay+3(a^2-b^2)t\bigr],\qquad q=b^2(y+6at),\qquad \rho=(b^3|t_0|)^{-1/2}=\left|b^3t-\frac{b^2(C_{22}-\beta C_{12})}{6\beta d}\right|^{-1/2}.
\]
For $s=(-1)^j$ ($j=1,2$), we introduce local coordinates $(\xi,\eta)$ around the leading lump positions via
\[
\begin{aligned}
t\to-\infty:\qquad &z=s\sqrt{12}\,\rho^{-1}+\xi,&\quad q=\eta,\\
t\to+\infty:\qquad &z=1+\xi,&\quad q=s\sqrt{12}\,\rho^{-1}+\eta.
\end{aligned}
\]
In terms of the spatial displacements $(\delta x,\delta y)$ from $(\xi,\eta)=(0,0)$ at fixed $t$, one has $\xi=b(\delta x+a\delta y)$ and $\eta=b^2\delta y$. In these coordinates, the fundamental lump solution \cite{qiu-mkp} is $bV_0(\xi,\eta)$, with
\[
V_0=4\operatorname{Im}\frac{G_\xi}{G}, \qquad G=\xi^2+\eta^2+\frac{\alpha^2}{S}+\frac{2i(\alpha\xi-\eta)}{S}.
\]
Expanding $f_{4,4}$ in Appendix~\ref{app:functions} as $\rho\to0$ yields, uniformly on compact subsets of the $(\xi,\eta)$-plane,
\[
\frac{\rho^2 f_{4,4}}{48(a^2+b^2)^2}=G+s\rho P_\pm+O(\rho^2), \qquad \frac{v}{b}=V_0+s\rho V_{1,\pm}+O(\rho^2),
\]
where the subscripts $-$ and $+$ correspond to $t\to-\infty$ and $t\to+\infty$, respectively, $V_{1,\pm}=4\operatorname{Im}\partial_\xi(P_\pm/G)$, and
\begin{align*}
\sqrt{12}\,P_-{}&=\xi^3+\xi\eta^2+\frac{3i\alpha\xi^2-2i\xi\eta-(2S-i\alpha)\eta^2}{S}\\
&+\frac{(2\alpha^4+\alpha^2+4i\alpha+1)\xi+2i\alpha(2\alpha-i)\eta}{S^2}+\frac{\alpha(2\alpha^3+2i\alpha^2+4\alpha+i)}{S^2},\\
\sqrt{12}\,P_+{}&=\eta^3+\eta\xi^2-\frac{3i\eta^2+2(S-i\alpha)\xi\eta+i\xi^2}{S}\\
&+\frac{(\alpha^4-2i\alpha^3+\alpha^2-6i\alpha-2)\eta+2i(\alpha^2-i\alpha-1)\xi}{S^2}-\frac{i\alpha(\alpha-2i)}{S^2}.
\end{align*}
Here, the uniform expansion of $v/b$ and its $(\xi,\eta)$-derivatives follows immediately from $\operatorname{Re}G\ge\alpha^2/S>0$.

At $(\xi,\eta)=(0,0)$, $V_0$ attains its strict global maximum $V_0(0,0)=8/\alpha$, with $\nabla_{\xi,\eta} V_0(0,0)=0$ and negative-definite Hessian
\[
\mathcal H
=16\begin{pmatrix}
-\dfrac{3\alpha^2+7}{\alpha^3}
&\dfrac{\alpha^2+5}{\alpha^4}\\[4pt]
\dfrac{\alpha^2+5}{\alpha^4}
&-\dfrac{\alpha^4+\alpha^2+4}{\alpha^5}
\end{pmatrix},
\qquad \det\mathcal H=\frac{768S^3}{\alpha^8}>0.
\]
Direct evaluation gives
\[
V_{1,-}(0,0)=-\frac{8}{\sqrt3\,\alpha}, \quad \nabla_{\xi,\eta} V_{1,-}(0,0)=\mathcal H\begin{pmatrix}\sqrt3/12\\0\end{pmatrix}, \quad V_{1,+}(0,0)=\frac{4}{\sqrt3\,\alpha^2},\quad \nabla_{\xi,\eta} V_{1,+}(0,0)=\begin{pmatrix}0\\0\end{pmatrix}.
\]
By the implicit function theorem applied to $\nabla_{\xi,\eta}(v/b)=0$, for sufficiently small $\rho$ there is a unique strict local maximum near $(0,0)$ at
\[
\begin{pmatrix}\xi\\\eta\end{pmatrix}
=-s\rho\mathcal H^{-1}\nabla_{\xi,\eta} V_{1,\pm}(0,0)+O(\rho^2),
\]
giving $(\xi,\eta)=(-\frac{s\sqrt3}{12}\rho,0)+O(\rho^2)$ as $t\to-\infty$ and $(\xi,\eta)=O(\rho^2)$ as $t\to+\infty$. Since $\nabla_{\xi,\eta} V_0(0,0)=0$, the $O(\rho)$ shift in $(\xi,\eta)$ affects the peak value only at $O(\rho^2)$, and thus
\[
\frac{A_j}{b}=\frac{8}{\alpha}+s\rho V_{1,\pm}(0,0)+O(\rho^2).
\]
Finally, $\rho=b^{-3/2}|t|^{-1/2}+O(|t|^{-3/2})$. Using $z=bZ$ and $q=b^2Y$ in \eqref{eq:coordinates}, we recover
\[
x=3(a^2+b^2)t+\frac{z}{b}-\frac{aq}{b^2},\qquad y=-6at+\frac{q}{b^2}.
\]
Substitution gives \eqref{eq:two-in} and \eqref{eq:two-out}. In particular, the outgoing center $z=1$ gives the $1/b$ offset in $x_j$. \qed
\end{proof}

\begin{figure}[ht]\centering
\panel{.325}{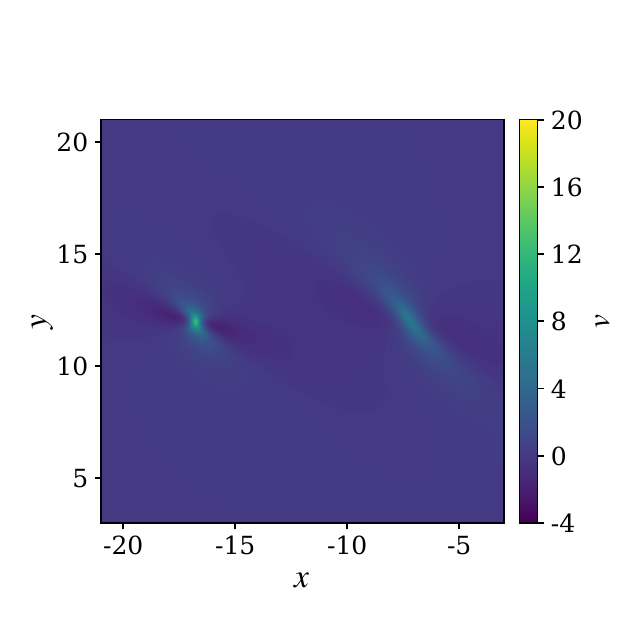}{(a) $t=-2$.}\hfill
\panel{.325}{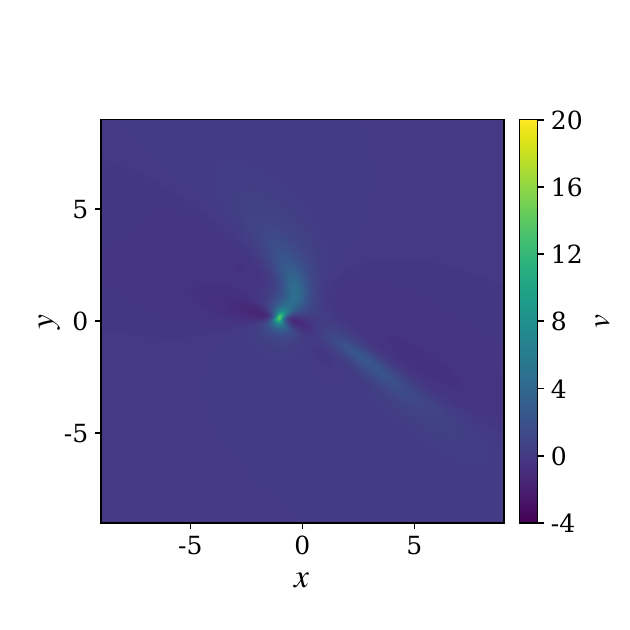}{(b) $t=0$.}\hfill
\panel{.325}{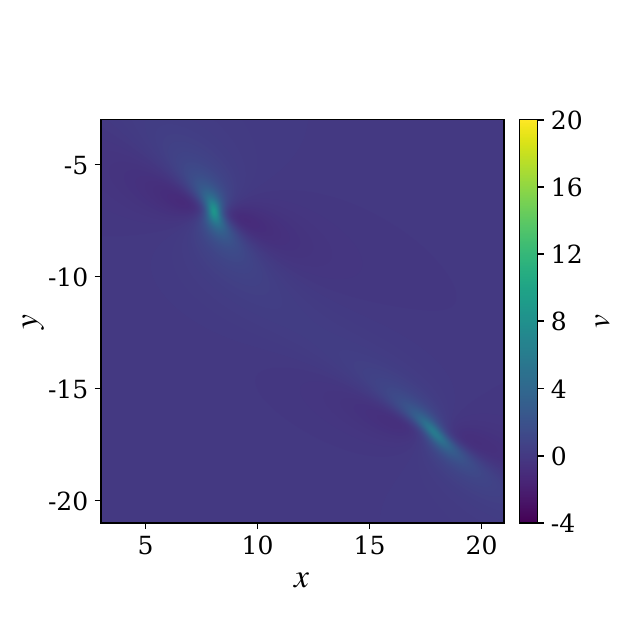}{(c) $t=2$.}
\caption{Second-order two-lump solution \eqref{eq:two} with $\lambda=1+i$, $\beta=d=1$, $C_{10}=1$, and $C_{20}=-3$.}\label{fig:two}
\end{figure}

The proofs of the remaining propositions below follow from analogous local expansions and stationary-point calculations and are omitted. Figure~\ref{fig:two} shows the evolution of the two-lump solution \eqref{eq:two}. For the comparison, let $f_{4,4}^{\mathrm{KP}}(Z,Y,t)$ denote the auxiliary function in Eq.~(A.2) of Ref.~\cite{qiu-arxiv}, with the co-moving coordinates \eqref{eq:coordinates}. This polynomial is even in $Y$. Since $u=2\partial_Z^2\ln f_{4,4}^{\mathrm{KP}}$ in the co-moving coordinates, we have
\[
 f_{4,4}^{\mathrm{KP}}(Z,-Y,t)=f_{4,4}^{\mathrm{KP}}(Z,Y,t),\qquad u(Z,-Y,t)=u(Z,Y,t).
\]
For all sufficiently large positive times, the reflection $(Z,Y)\mapsto(Z,-Y)$ preserves strict local maxima and their amplitudes. Together with Proposition~\ref{prop:two}, this gives
\[
 A_{\max}^{\mathrm{mKP}}-A_{\min}^{\mathrm{mKP}}=\frac{8b^{3/2}}{\sqrt3\,a^2}t^{-1/2}+O(t^{-1}), \qquad A_{\max}^{\mathrm{KP}}-A_{\min}^{\mathrm{KP}}=0, \qquad t\to+\infty.
\]
Thus, as $t\to+\infty$, the two mKP-I lumps differ in amplitude at order $t^{-1/2}$, although both amplitudes converge to the same limit $8b^2/a$. By contrast, as $t\to-\infty$, both equations have $|t|^{-1/2}$ amplitude corrections of opposite signs.

\subsection{Anomalous scattering of the second-order three-lump solution}\label{sec:old}

Canceling the Taylor coefficients through order five yields
\[
 d=0,\qquad C_{10}=-\frac{\beta+1}{\beta-1},\qquad C_{20}=\frac{\beta+1}{\beta-1},\qquad C_{22}=\beta C_{12}, \qquad \beta\notin\{0,1,-1\}.
\]
The lump solution is
\begin{equation}\label{eq:oldthree}
 v=4\operatorname{Im}\frac{(f_{4,6})_x}{f_{4,6}},
\end{equation}
where $f_{4,6}$ is independent of $C_{12}$, and $C_{13}$ and $C_{23}$ enter only through the time shift
\[
 t \mapsto t-\frac{C_{23}-\beta C_{13}}{\beta(\beta^2-1)}.
\]
When the higher-order phase coefficients vanish, this lump solution reduces to the second-order three-lump solution reported in \cite{qiu-mkp}. The following proposition gives its asymptotics as $t\to\pm\infty$. 

\begin{proposition}\label{prop:cubic-three}
Take $C_{23}-\beta C_{13}=0$ and set $\theta_j=2(j-1)\pi/3$, $j=1,2,3$. As $t\to\pm\infty$, the positions and amplitudes of the three lumps in \eqref{eq:oldthree} satisfy
\begin{equation*}
 \begin{aligned}
 &x_j=3(a^2+b^2)t+\frac{b\cos\theta_j-a\sin\theta_j}{b^2}\left(\sqrt[3]{-12b^3t}-\frac1{\sqrt[3]{-12b^3t}}\right)+O(|t|^{-2/3}),\\
 &y_j=-6at+\frac{\sin\theta_j}{b^2}\left(\sqrt[3]{-12b^3t}-\frac1{\sqrt[3]{-12b^3t}}\right)+O(|t|^{-2/3}),\\
 &A_j=\frac{8b^2}{a}+
 \frac{8b^2[a(a^2+4b^2)\cos\theta_j-b(a^2-2b^2)\sin\theta_j]}
 {a^2(a^2+b^2)(12b^3|t|)^{2/3}}+O(|t|^{-1}).
 \end{aligned}
\end{equation*}
\end{proposition}

\section{Third-order lump solutions}\label{sec:third}

\subsection{Anomalous scattering of the third-order three-lump solution}\label{sec:thirdthree}

\begin{figure}[ht]\centering
\panel{.325}{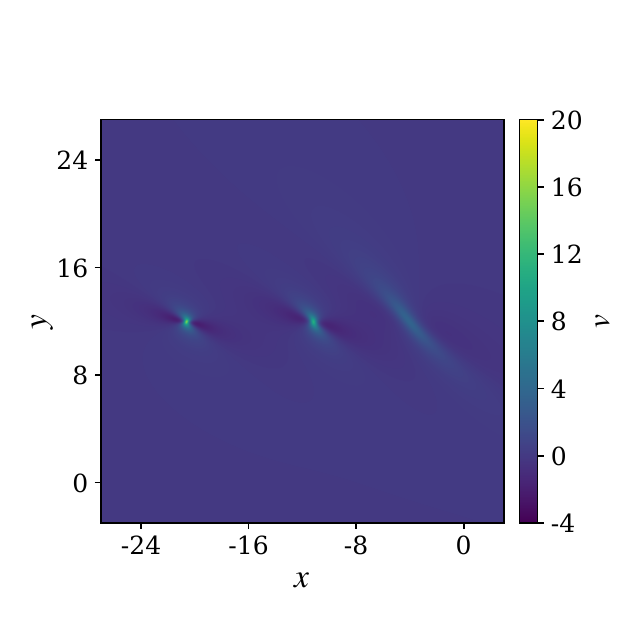}{(a) $t=-2$.}\hfill
\panel{.325}{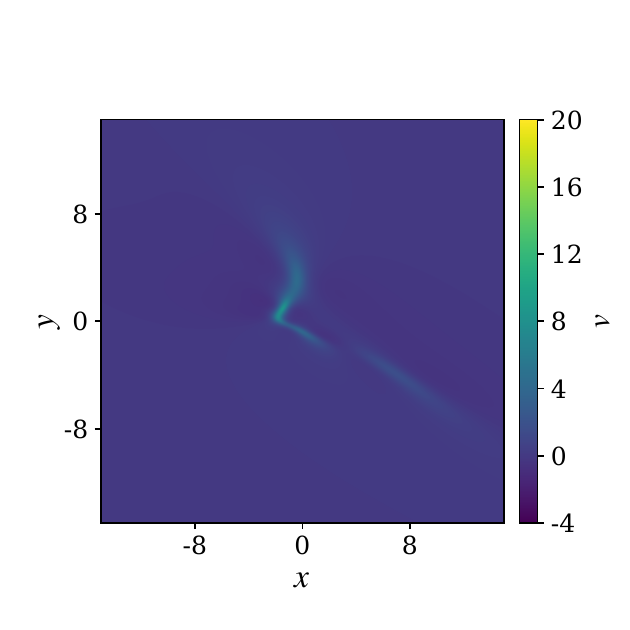}{(b) $t=0$.}\hfill
\panel{.325}{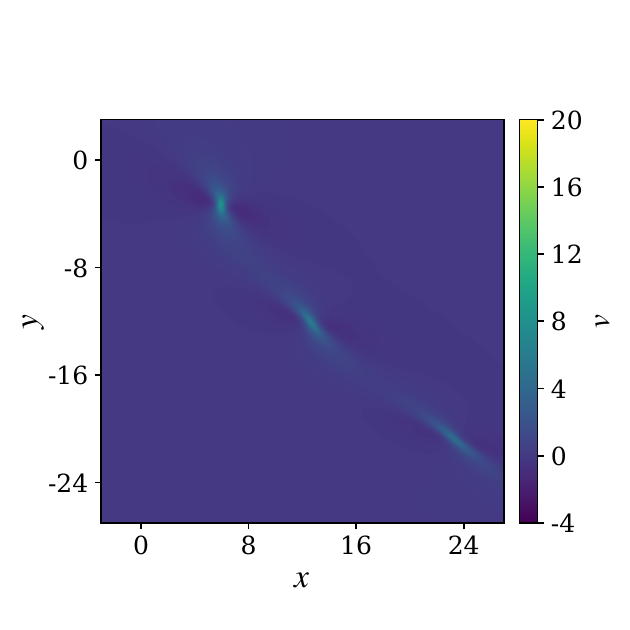}{(c) $t=2$.}
\caption{Third-order three-lump solution \eqref{eq:thirdthree} with $\lambda=1+i$, $(\beta_j)=(1,1,1)$, $(d_j)=(0,1,4)$, and $(C_{j0})=(1/2,-1,-5/2)$.}\label{fig:three}
\end{figure}

For $M=3$, setting
\[
 (\beta_1,\beta_2,\beta_3)=(1,1,1),\quad (d_1,d_2,d_3)=(0,1,4),\quad (C_{10},C_{20},C_{30})=(1/2,-1,-5/2), \quad C_{12}=(4C_{22}-C_{32})/3
\]
cancels all Taylor coefficients below order six and yields the third-order three-lump solution (see Fig.~\ref{fig:three})
\begin{equation}\label{eq:thirdthree}
 v=4\operatorname{Im}\frac{(f_{6,6})_x}{f_{6,6}}.
\end{equation}

\begin{proposition}\label{prop:thirdthree}
Take $C_{22}=C_{32}$ and $3C_{13}-4C_{23}+C_{33}=0$. For $j=1,2$, set
\[
 (m_{1,j},m_{2,j},c_j)=
 \begin{cases}
 (6(-1)^j,0,-1/3),&t<0,\\
 (0,6(-1)^j,8/3),&t>0.
 \end{cases}
\]
As $t\to\pm\infty$, the positions and amplitudes of the two outer lumps in \eqref{eq:thirdthree} satisfy
\begin{equation*}
 \begin{aligned}
 &x_j=3(a^2+b^2)t+\frac{bm_{1,j}-am_{2,j}}{b^2}\sqrt{b^3|t|}+\frac{c_j}b+O(|t|^{-1/2}),\\
 &y_j=-6at+\frac{m_{2,j}}{b^2}\sqrt{b^3|t|}+O(|t|^{-1/2}),\\
 &A_j=\frac{8b^2}{a}+\frac{\operatorname{sgn}t}{\sqrt{b^3|t|}}
 \left(\frac{4b^2m_{1,j}}{3a}+\frac{2b^3m_{2,j}}{3a^2}\right)+O(|t|^{-1}).
 \end{aligned}
\end{equation*}
The position and amplitude of the central lump satisfy
\begin{equation*}
 \begin{aligned}
 &x_3=3(a^2+b^2)t+\frac2{3b}-\frac{56}{243b^4t}+O(|t|^{-2}),\quad y_3=-6at+O(|t|^{-2}),\\
 &A_3=\frac{8b^2}{a}-\frac{4(7a^2+16b^2)}{9ab(a^2+b^2)t}+O(|t|^{-2}).
 \end{aligned}
\end{equation*}
\end{proposition}

\subsection{Anomalous scattering of the third-order four-lump solution}\label{sec:four}

\begin{figure}[ht]\centering
\panel{.325}{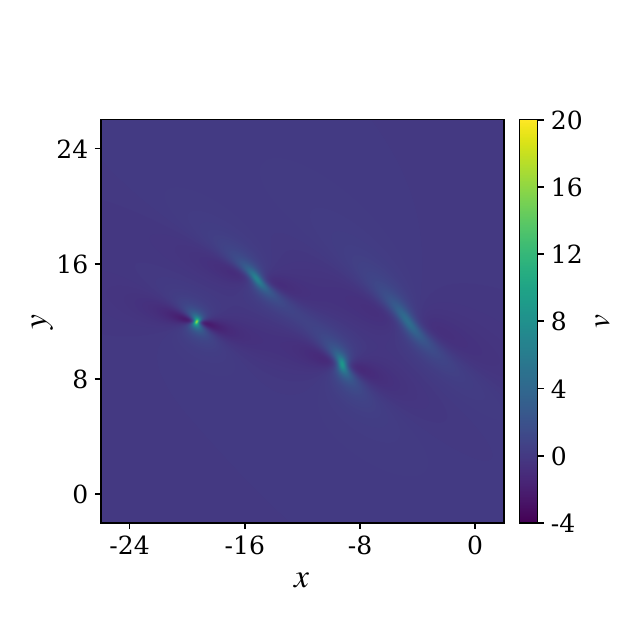}{(a) $t=-2$.}\hfill
\panel{.325}{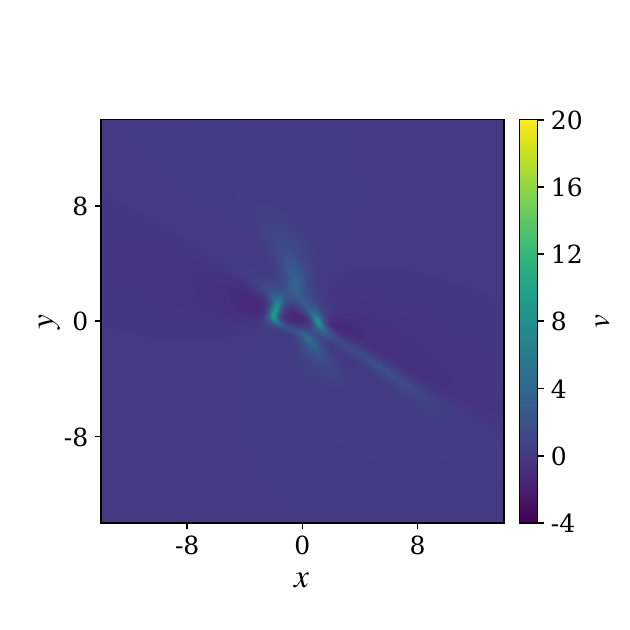}{(b) $t=0$.}\hfill
\panel{.325}{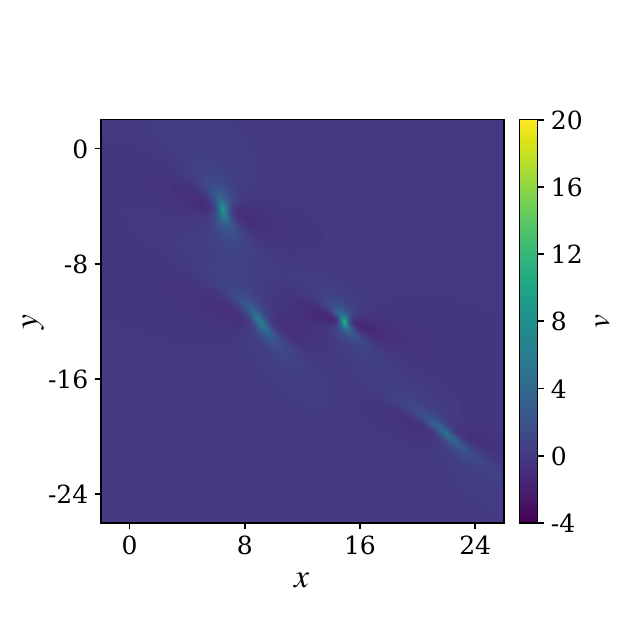}{(c) $t=2$.}
\caption{Third-order four-lump solution \eqref{eq:four} with $\lambda=1+i$ and the parameters in \eqref{eq:four-data}.}\label{fig:four}
\end{figure}

We take
\begin{equation}\label{eq:four-data}
 \begin{aligned}
 &(\beta_1,\beta_2,\beta_3)=(1,8,11),\qquad(d_1,d_2,d_3)=(0,1,2),\\
 &(C_{10},C_{20},C_{30})=(-5/3,15,-35/2),\qquad C_{js}=0\ (s\ge2).
 \end{aligned}
\end{equation}
With this choice, the Taylor coefficients through order seven vanish, leading to the four-lump solution shown in Fig.~\ref{fig:four},
\begin{equation}\label{eq:four}
 v=4\operatorname{Im}\frac{(f_{6,8})_x}{f_{6,8}}.
\end{equation}

\begin{proposition}\label{prop:four}
For $j=1,2,3,4$, define
\[
 (m_{1,j},m_{2,j},c_j)=
 \begin{cases}
 \bigl((-1)^j\sqrt{12(\sqrt2-\operatorname{sgn}t)},0,0\bigr),&j=1,2,\\
 \bigl(0,(-1)^j\sqrt{12(\sqrt2+\operatorname{sgn}t)},1+\sqrt2\operatorname{sgn}t\bigr),&j=3,4.
 \end{cases}
\]
As $t\to\pm\infty$, the positions and amplitudes of the four lumps in \eqref{eq:four} satisfy
\begin{equation*}
 \begin{aligned}
 &x_j=3(a^2+b^2)t+\frac{bm_{1,j}-am_{2,j}}{b^2}\sqrt{b^3|t|}+\frac{c_j}b+O(|t|^{-1/2}),\\
 &y_j=-6at+\frac{m_{2,j}}{b^2}\sqrt{b^3|t|}+O(|t|^{-1/2}),\\
 &A_j=\frac{8b^2}{a}+\frac{\operatorname{sgn}t}{\sqrt{b^3|t|}}
 \left(\frac{4b^2m_{1,j}}{3a}+\frac{2b^3m_{2,j}}{3a^2}\right)+O(|t|^{-1}).
 \end{aligned}
\end{equation*}
\end{proposition}

\subsection{Anomalous scattering of the third-order five-lump solution}\label{sec:five}

\begin{figure}[ht]\centering
\panel{.325}{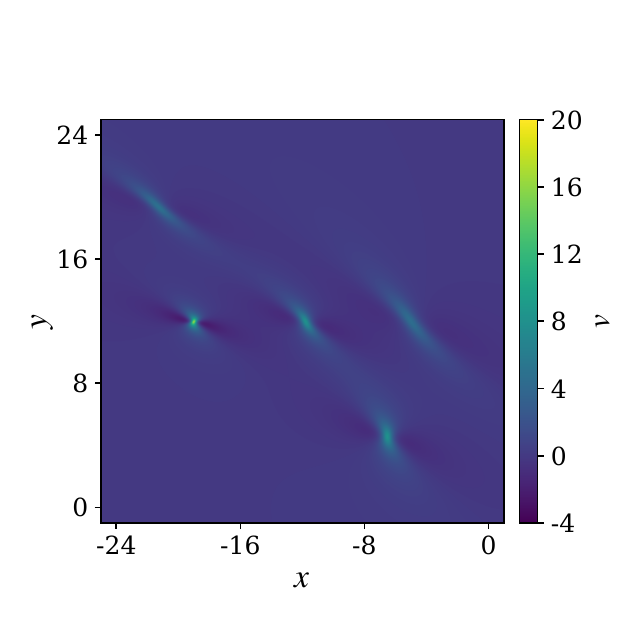}{(a) $t=-2$.}\hfill
\panel{.325}{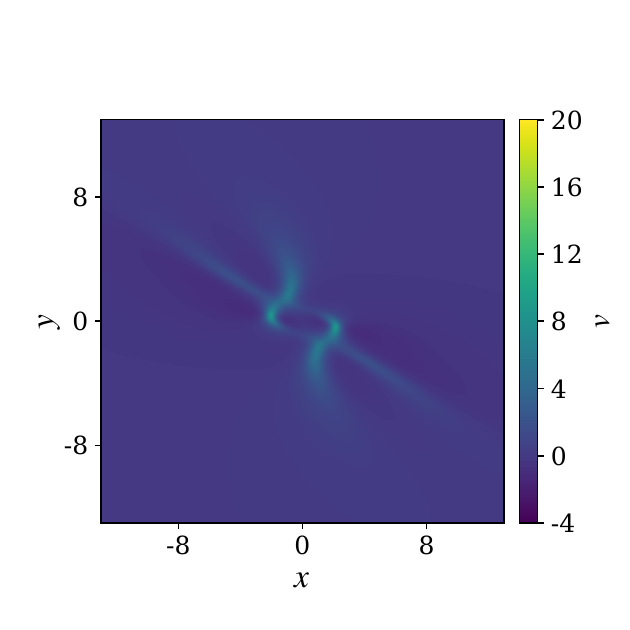}{(b) $t=0$.}\hfill
\panel{.325}{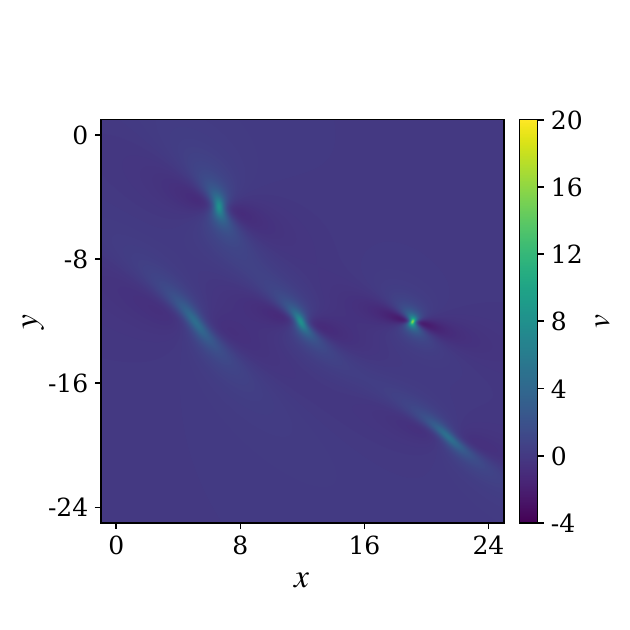}{(c) $t=2$.}
\caption{Third-order five-lump solution \eqref{eq:five} with $\lambda=1+i$ and the parameters in \eqref{eq:five-data}.}\label{fig:five}
\end{figure}

We choose
\begin{equation}\label{eq:five-data}
 \begin{aligned}
 &(\beta_1,\beta_2,\beta_3)=(1,1,7),\qquad(d_1,d_2,d_3)=(0,8,4),\\
 &(C_{10},C_{20},C_{30})=(-3/2,-3/2,-12/5),\qquad C_{js}=0\ (s\ge2).
 \end{aligned}
\end{equation}
Here, the first nonvanishing Taylor coefficient occurs at order $\varepsilon^{10}$, giving the five-lump solution (illustrated in Fig.~\ref{fig:five})
\begin{equation}\label{eq:five}
 v=4\operatorname{Im}\frac{(f_{6,10})_x}{f_{6,10}}.
\end{equation}

\begin{proposition}\label{prop:five}
For the four outer lumps ($j=1,2,3,4$), define
\[
 (m_{1,j},m_{2,j},c_j)=
 \begin{cases}
 \bigl((-1)^j\sqrt{12\sqrt5},0,0\bigr),&j=1,2,\\
 \bigl(0,(-1)^j\sqrt{12\sqrt5},\sqrt5\operatorname{sgn}t\bigr),&j=3,4.
 \end{cases}
\]
As $t\to\pm\infty$, the positions and amplitudes of these lumps satisfy
\begin{equation*}
 \begin{aligned}
 &x_j=3(a^2+b^2)t+\frac{bm_{1,j}-am_{2,j}}{b^2}\sqrt{b^3|t|}+\frac{c_j}b+O(|t|^{-1/2}),\\
 &y_j=-6at+\frac{m_{2,j}}{b^2}\sqrt{b^3|t|}+O(|t|^{-1/2}),\\
 &A_j=\frac{8b^2}{a}+\frac{\operatorname{sgn}t}{\sqrt{b^3|t|}}
 \left(\frac{4b^2m_{1,j}}{3a}+\frac{2b^3m_{2,j}}{3a^2}\right)+O(|t|^{-1}).
 \end{aligned}
\end{equation*}
The central lump satisfies
\begin{equation*}
 \begin{aligned}
 &x_5=3(a^2+b^2)t-\frac1{3b^4t}+O(|t|^{-2}),\quad
 y_5=-6at+O(|t|^{-2}),\\
 &A_5=\frac{8b^2}{a}+\frac{2(2a^6+5a^4b^2-13a^2b^4+20b^6)}{9ab^4(a^2+b^2)^3t^2}+O(|t|^{-3}).
 \end{aligned}
\end{equation*}
\end{proposition}

\subsection{Anomalous scattering of the third-order six-lump solution}\label{sec:six}

\begin{figure}[ht]\centering
\panel{.325}{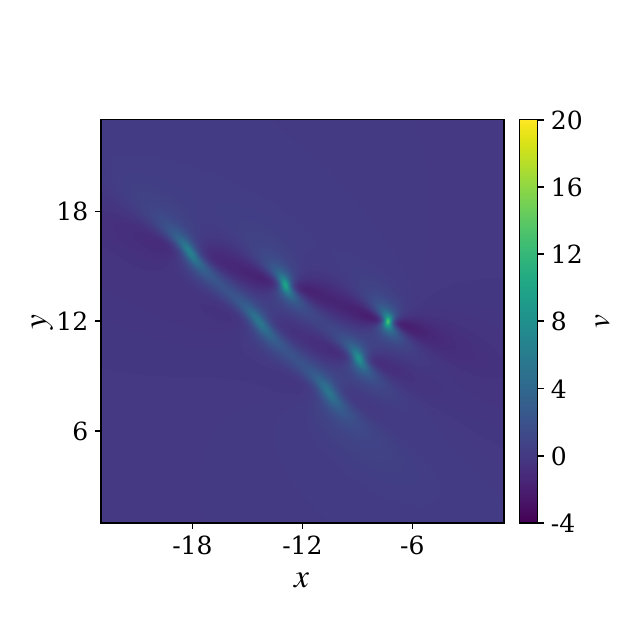}{(a) $t=-2$.}\hfill
\panel{.325}{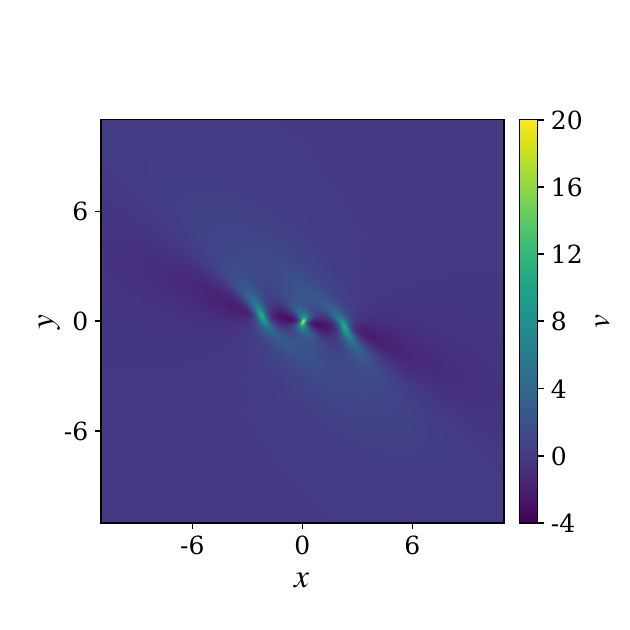}{(b) $t=0$.}\hfill
\panel{.325}{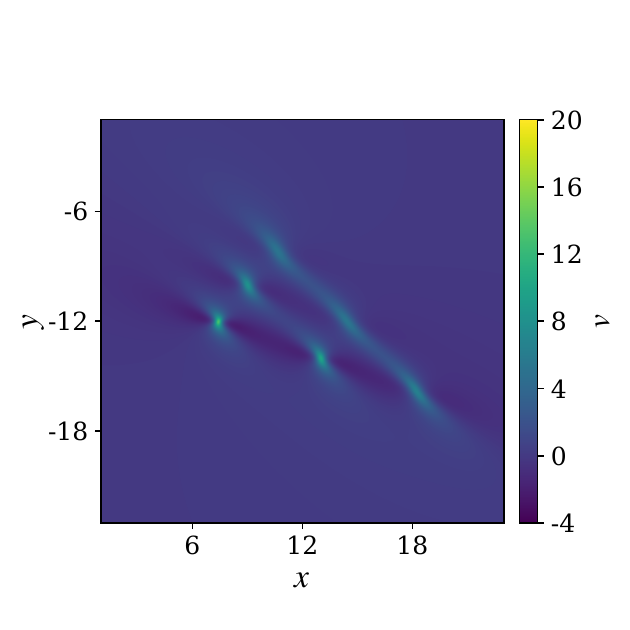}{(c) $t=2$.}
\caption{Third-order six-lump solution \eqref{eq:six} with $\lambda=1+i$, the parameters in \eqref{eq:six-data}, $T=t$, and $H=0$.}\label{fig:six}
\end{figure}

Finally, for the parameter choice
\begin{equation}\label{eq:six-data}
 \begin{aligned}
 &(\beta_1,\beta_2,\beta_3)=(1,2,3),\quad d_1=d_2=d_3=0,\quad(C_{10},C_{20},C_{30})=(-6,15,-10),\\
 &C_{12}=C_{32}/3,\quad C_{22}=2C_{32}/3,\quad C_{13}=(4C_{23}-C_{33})/5,\quad C_{14}=(4C_{24}-C_{34})/5.
 \end{aligned}
\end{equation}
the leading nonzero term at order $\varepsilon^{12}$ gives the six-lump solution (see Fig.~\ref{fig:six})
\begin{equation}\label{eq:six}
 v=4\operatorname{Im}\frac{(f_{6,12})_x}{f_{6,12}},
\end{equation}
in which the higher-order phases enter through the effective parameters
\[
 T=t+\frac{3C_{23}-2C_{33}}{30},\qquad H=5C_{15}-4C_{25}+C_{35}.
\]

\begin{proposition}\label{prop:six}
Keep all phase parameters fixed and set $\theta_j=2(j-1)\pi/3$, $j=1,2,3$. The six lumps form two groups, labeled by $(+,j)$ and $(-,j)$. As $t\to\pm\infty$, their positions and amplitudes satisfy
\begin{equation*}
 \begin{aligned}
 &x_{\pm,j}=3(a^2+b^2)t+\frac{b\cos\theta_j-a\sin\theta_j}{b^2}\left(\sqrt[3]{-(30\pm18\sqrt5)b^3T}-\frac{5\pm\sqrt5}{2\sqrt[3]{-(30\pm18\sqrt5)b^3T}}\right)+O(|t|^{-2/3}),\\
 &y_{\pm,j}=-6at+\frac{\sin\theta_j}{b^2}\left(\sqrt[3]{-(30\pm18\sqrt5)b^3T}-\frac{5\pm\sqrt5}{2\sqrt[3]{-(30\pm18\sqrt5)b^3T}}\right)+O(|t|^{-2/3}),\\
 &A_{\pm,j}=\frac{8b^2}{a}+\frac{4b^2(5\pm3\sqrt5)\{a(a^2+4b^2)\cos\theta_j-b(a^2-2b^2)\sin\theta_j\}}{a^2(a^2+b^2)\bigl[\sqrt[3]{-(30\pm18\sqrt5)b^3T}\bigr]^2}+O(|t|^{-1}).
 \end{aligned}
\end{equation*}
\end{proposition}

\section{Discussion and conclusions}\label{sec:conclusion}

In this letter, we have constructed a second-order two-lump solution and third-order three-, four-, five-, and six-lump solutions of the mKP-I equation using the GLWL method \cite{qiu-arxiv}, while also recovering the second-order three-lump solution obtained in \cite{qiu-mkp}. Their asymptotics describe anomalous scattering with the same leading separation scales as the corresponding KP-I solutions. For the second-order two-lump solution, the outgoing mKP-I lump amplitudes differ at order $t^{-1/2}$, whereas the corresponding KP-I lump amplitudes are exactly equal for all sufficiently large positive times. The same contrast with the KP-I solutions \cite{qiu-arxiv} occurs for the outgoing outer lump pair of the third-order three-lump solution and the outgoing $j=3,4$ lump pairs of the third-order four- and five-lump solutions.

The regularity of general degenerate lump solutions of the mKP-I equation remains an open problem. We have verified that every solution plotted in Figs.~\ref{fig:two}--\ref{fig:six} is nonsingular for all real $(x,y,t)$. For general $\lambda$, deriving necessary and sufficient regularity conditions may require the KP reduction method and an analysis in terms of Schur polynomials \cite{yjkkp}. We leave this problem for future work.

Another natural direction is to apply this construction to other $(2+1)$-dimensional integrable equations. Since the classical long-wave limit method applies to a broad range of integrable systems \cite{ablo-lump}, we hope to extend the GLWL method to the Davey--Stewartson equations and related models.

\medskip
\noindent \textbf{Acknowledgements} This work is supported by the Kaiyuan International Mathematical Sciences Institute, the Fundamental Research Funds for the Central Universities (JKF2025077492962), and the Academic Excellence Foundation of BUAA for Ph.D. Students.

\section*{Declaration of generative AI use}
During the preparation of this manuscript, the authors used OpenAI GPT-6.1 Sol and GPT-6 Astra to assist with mathematical derivations. They reviewed, verified, and edited all AI-generated content and take full responsibility for its correctness and for the final manuscript.

\section*{Declarations}
\noindent \textbf{Conflict of interest} The authors declare that they have no conflict of interest.

\noindent\textbf{Data availability} The computational materials supporting this work are not publicly available. Requests for further information should be addressed to the authors.

\appendix
\setcounter{equation}{0}
\renewcommand{\theequation}{A.\arabic{equation}}
\section{Auxiliary function}\label{app:functions}
We use the co-moving coordinates
\begin{equation}\label{eq:coordinates}
 X=x-3(a^2+b^2)t,\qquad Y=y+6at,\qquad Z=X+aY.
\end{equation}
The auxiliary function for the lump solution in Eq.~\eqref{eq:two} is
%\begingroup\small
\begin{align*}
f_{4,4}&=\bigl(a^{2} + b^{2}\bigr)^{2}\Bigl[Z^{4} b^{4} +2 Z^{2} b^{2} \bigl(Y^{2} b^{4} + 12 b^{3} t_0 + 2\bigr)+8 Z b \bigl(- Y^{2} b^{4} + 1\bigr) +8 +Y^{4} b^{8} +8 Y^{2} b^{4} +144 b^{6} t_0^{2}-24 Y^{2} b^{7} t_0\Bigr]\\
&+2 b \bigl(a^{2} + b^{2}\bigr)\Bigl[2 i Z^{3} a b^{3} -Z^{2} b^{3} \bigl(2 i Y b^{2} + 3\bigr)+2 i Z b \bigl(Y^{2} a b^{4} + 4 Y b^{3} + 12 a b^{3} t_0 + 2 a\bigr) -2 b -Y^{2} b^{5}\\
&\qquad-12 b^{4} t_0 +4 i a -8 i Y b^{3} -2 i Y^{3} b^{7} -4 i Y^{2} a b^{4}+24 i Y b^{6} t_0\Bigr]\\
&+b^{4}\Bigl[4 Z^{2} b \bigl(2 i a + b\bigr) +4 Z \bigl(2 Y a b^{2} - 4 i Y b^{3} - i a - 2 b\bigr) -3 -4 Y^{2} b^{4}+48 b^{3} t_0 -16 Y a b +20 i Y b^{2} -8 i Y^{2} a b^{3}+96 i a b^{2} t_0\Bigr],
\end{align*}
%\endgroup
where $t_0=t-(C_{22}-\beta C_{12})/(6\beta db)$.

\end{sloppypar}
\end{document}